\documentclass{eptcs}
\providecommand{\event}{ThEdu 2020 Postproceedings} 
\providecommand{\thisvolume}[1]{this volume of EPTCS, Open Publishing Association}

\usepackage{underscore} 
\usepackage{paralist}   
\usepackage{graphicx}   
\usepackage{amsmath}    

\title{Lucas-Interpretation on Isabelle's Functions}
\author{Walther Neuper
  \institute{Johannes Kepler University, Linz, Austria}
  \email{walther.neuper@jku.at}
}
\def\titlerunning{Lucas-Interpretation on Isabelle's Functions}
\def\authorrunning{W. Neuper}

\def\isac{\textit{ISAC}}
\def\sisac{\textit{ISAC}}
\def\LI{\textit{LI}}

\begin{document}
\maketitle

\begin{abstract}
Software tools of Automated Reasoning are too sophisticated for general use in mathematics education and respective reasoning, while Lucas-Interpretation provides a general concept for integrating such tools into educational software with the purpose to reliably and flexibly check formal input of students.

This paper gives the first technically concise description of Lucas-Interpretation at the occasion of migrating a prototype implementation to the function package of the proof assistant Isabelle. The description shows straightforward adaptations of Isabelle's programming language and shows, how simple migration of the interpreter was, since the design (before the function package has been introduced to Isabelle) recognised appropriateness of Isabelle's terms as middle end.

The paper gives links into the code in an open repository as invitation to readers for re-using the prototyped code or adopt the general concept. And since the prototype has been designed before the function package was implemented, the paper is an opportunity for recording lessons learned from Isabelle's development of code structure.
\end{abstract}

\section{Introduction}\label{sec:intro}
This paper concerns application of Automated Reasoning (AR) to education, the issue to narrow the gap between powerful, but highly sophisticated technologies of AR on the one side and requirements of education in mathematics as taught in engineering studies. Lucas-Interpretation (in the sequel abbreviated by \LI) provides a general concept for integrating AR tools into educational software with the purpose to reliably and flexibly check formal input of students.

A prerequisite for \LI{} is a logical framework, which models rigorous formal derivation, in particular forward reasoning. There are several educational software products of this kind: \cite{tBaBoEr07a} calls ``structuredderivation'' what here is called ``calculation'' and uses the PVS prover in the background, IMPS is an Interactive Mathematical Proof System intended to provide organizational and computational support for the traditional techniques of mathematical reasoning~\cite{Farmer-al:93}, and in geometry software AR tools are used~\cite{ggb:atp-15,gclc-06}, too. But all the above mentioned systems are poor in guided interaction.

A logic course for freshmen in AI and Computer Science\footnote{\url{http://fmv.jku.at/logic/}} accompanies all content with software. And corresponding experience shows~\cite{RISC5885} again, that much of the software lacks interactivity and feedback sufficient for independent learning\footnote{Currently independent learning is best addressed by ``flipped classes''~\cite{flip-class-meta}} in particular in application to mathematics.

\medskip
In principle~\cite{EPTCS290.6}, AR provides the most powerful tools for independent learning in formal mathematics, captured by naive requirements like these: given a formal problem specification, AR checks steps of forward reasoning towards a problem solution, a step determined by an input term or by an input theorem to be applied to the current state of solution; and the system can provide a next step, if a student gets stuck. Nowadays such naive requirements might be reconsidered, since proof assistants like Isabelle~\cite{Isa-tutor08} hide the intricacies of AR tools and present notation to the user close to standard mathematics.

\LI{} aims at meeting these requirements for more than a decade, when Peter Lucas\footnote{\url{https://austria-forum.org/af/AustriaWiki/Peter_Lucas_(Informatiker)}} shifted his interests from programming languages~\cite{pl:formal-lang-hist} to education. His specific contribution has been named after him, and now reveals the ingenuity of the original design, when the prototype has been migrated: a  proprietary version~\cite{wn:proto-sys} was migrated to Isabelle's function  package~\cite{funpack-tutorial,krauss} with surprising little effort.

\medskip
Herewith the first technically concise description of \LI{} is given, after theoretical considerations~\cite{wn:lucas-interp-12} and application oriented discussions~\cite{wneuper:gcd-coimbra,thedu16:lucin-user-view,wn:lucin-thedu18}. Since \LI{} was implemented in the \sisac-project\footnote{\url{https://isac.miraheze.org/wiki/History}}  several years before the function package was introduced to the proof assistant Isabelle, we take the paper as an opportunity for recording lessons learned from Isabelle's development.

\paragraph{The paper is structured as follows:} \S\ref{lucin} introduces \LI{}, the concept in \S\ref{concept-LI} illustrated with examples in \S\ref{expl-lucin}. \S\ref{edu} notes \LI's relevance for educational mathematics software. Adaptation of the  programs for \LI{} to Isabelle's function package is described in \S\ref{lucin-funpack}, where \S\ref{tactics} describes the tactics, \S\ref{tacticals} the tacticals and \S\ref{prog-expr} briefly describes evaluation by rewriting. The parse-tree generated by the function package is interpreted by \LI{} as shown in \S\ref{LI-impl}; \S\ref{scanning} gives a description of how the parse-tree is scanned, \S\ref{ctxt} of how \LI{} uses Isabelle's \textit{Proof.context} and \S\ref{embedding} of how the prototype's mathematics-engine embeds and guards \LI{}. Lessons have been learned from Isabelle's code structure and development process \S\ref{learned-devel}, from specific Isabelle features (\S\ref{learned-feature}) and now direct future prototyping (\S\ref{learned-isac}). In \S\ref{summary-concl} the summary concludes with expectations on \LI{} widely applied to education in mathematics.

\section{Lucas--Interpretation (\LI)}\label{lucin}
The interpreter is named after the inventor of top-down-parsing in the ALGOL project~\cite{pl:formal-lang-hist}, Peter Lucas. As a dedicated expert in programming languages he initially objected ``yet another programming language'' (in analogy to ``Yacc'' \cite{Yacc-1975}), but then he helped to clarify the unusual requirements for a novel programming language in the \sisac-project, which later led to the notion of \LI. 

\LI{} is the most prominent component in a prototype developed in the \sisac-project, there embedded in a mathematics-engine, which interacts with a dialogue-module in a Java-based front-end managing interaction with students (briefly touched below in \S\ref{edu}).

\subsection{The Concept of \LI}\label{concept-LI}
The concept of \LI{} is simple: \LI{} \emph{acts as a debugger on functional programs with hard-coded breakpoints, where control is handed over to a student; a student, however, does \emph{not} interact with the software presenting the program, but with a software presenting steps of forward reasoning, where the steps are rigorously constructed by tactics acting as the break-points mentioned}. Following the \LI{} terminology, we will call ``programs'' the functions defined with the function package and refer to their evaluation as ``execution''.

\paragraph{Types occurring in the signatures} of \LI{} are as follows. Besides the program of type \texttt{\small Program.T} there is an interpreter state \texttt{\small Istate.T}, a record type passing data from one step of execution to the next step, in particular a location in the program, where the next step will be read off, and an environment for evaluating the step.
As invisible in the program language as the interpreter state is \texttt{\small Calc.T}, a ``calculation'' as a sequence of steps in forward reasoning, a variant of ``structured derivations'' \cite{back-SD-2010}. Visible in the language and in the signature, however, are the tactics \texttt{\small Tactic.T}, which create steps in a calculation.

Novel in connection with calculations is the idea to maintain a logical context in order to provide automated provers with data required for checking input by the student. Isabelle's \texttt{\small Proof.context}, as is, turned out perfect for this task~\cite{mlehnf:bakk-11}.

\paragraph{The signatures of the main functions,} the functions \texttt{\small find\_next\_step}, \texttt{\small locate\_input\_tactic} and \texttt{\small locate\_input\_term}, are as follows:\label{sig-lucin}
\begin{verbatim}
  signature LUCAS_INTERPRETER =
  sig
    datatype next_step_result =
        Next_Step of Istate.T * Proof.context * Tactic.T
      | Helpless | End_Program of Istate.T * Tactic.T
    val find_next_step: Program.T -> Calc.T -> Istate.T -> Proof.context
      -> next_step_result
  
    datatype input_tactic_result =
        Safe_Step of Istate.T * Proof.context * Tactic.T
      | Unsafe_Step of Istate.T * Proof.context * Tactic.T
      | Not_Locatable of message
    val locate_input_tactic: Program.T -> Calc.T -> Istate.T -> Proof.context
        -> Tactic.T -> input_tactic_result
  
    datatype input_term_result =
      Found_Step of Calc.T | Not_Derivable
    val locate_input_term: Calc.T -> term -> input_term_result
  end
\end{verbatim}
\begin{description}
\item[\texttt{\small find\_next\_step}] accomplishes what usually is expected from a \texttt{\small Program.T}: find a \texttt{\small Next\_Step} to be executed, in the case of \LI{} to be inserted into a \texttt{\small Calc.T} under construction. This step can be a \texttt{\small Tactic.T}, directly found in \texttt{\small next\_step\_result}, or a \texttt{\small term} produced by applying the tactic. If such a step cannot be found (due to previous student interaction), \LI{} is \texttt{\small Helpless}.

\item[\texttt{\small locate\_input\_tactic}] gets a  \texttt{\small Tactic.T} as argument (which has been input and checked applicable in  \texttt{\small Calc.T}) and tries to locate it in the \texttt{\small Prog.T} such that a \texttt{\small Next\_Step} can be generated from the subsequent location in the \texttt{\small Program.T}. A step can be an \texttt{\small Unsafe\_Step}, if the input \texttt{\small Tactic.T} cannot safely be associated with any tactic in the \texttt{\small Program.T}. This function has a signature similar to \texttt{\small find\_next\_step} (here the respective input \texttt{\small Tactic.T} and the one in the result are the same) in order to unify internal technicalities of \LI{}.

\item[\texttt{\small locate\_input\_term}] tries to find a derivation from the current \texttt{\small Proof.context} for the term input to \texttt{\small Calc.T} and to locate a \texttt{\small Tactic.T} in the \texttt{\small Program.T}, which as \texttt{\small Found\_Step} allows to find a next step later. Such a \texttt{\small Found\_Step} can be located in another program (a sub-program or a calling program); thus \texttt{\small Program.T}, \texttt{\small Istate.T} and \texttt{\small Proof.context} are packed into \texttt{\small Calc.T} at appropriate positions and do not show up in the signature.
\end{description}

AR is concern of the third function \texttt{\small locate\_input\_term}, actually a typical application case for Isabelle's Sledgehammer~
but not yet realised and preliminarily substituted by \sisac's simplifier. \texttt{\small locate\_input\_term} is the function used predominantly in interaction with students: these input term by term into the calculation under construction (as familiar from paper\&pencil work) and \LI{} checks for correctness automatically.

\subsection{Examples for \LI}\label{expl-lucin}
\LI's novel concept relating program execution with construction of calculations is best demonstrated by examples. The first one is from structural engineering given by the following problem statement:\\
\medskip

\begin{minipage}{0.55\textwidth}\label{bend-line-expl}
\textit{Determine the bending line of a beam of length $L$,
which consists of homogeneous material, which is clamped on one side
and which is under constant line load $q_0$; see the Figure right.}\\
\end{minipage}
\hfill
\begin{minipage}{0.40\textwidth}
  \includegraphics[width=0.6\textwidth]{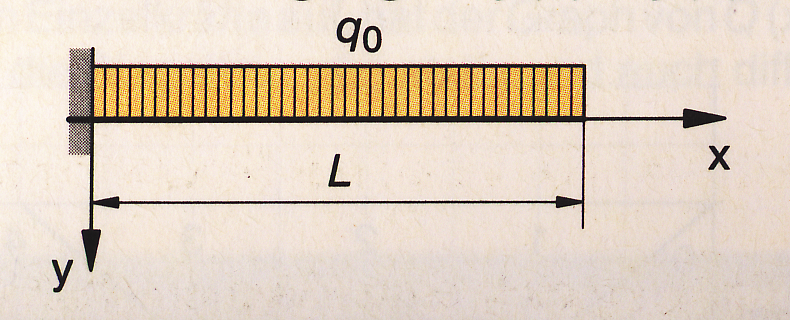}\\
\end{minipage}

\medskip
This problem is solved by the above program, which is shown by a screen-shot\footnote{The corresponding code is at \url{https://hg.risc.uni-linz.ac.at/wneuper/isa/file/df1b56b0d2a2/src/Tools/isac/Knowledge/Biegelinie.thy\#l174}.
}
This problem is solved by the above program, which is shown by a screen-shot\footnote{The corresponding code is at \url{https://hg.risc.uni-linz.ac.at/wneuper/isa/file/df1b56b0d2a2/src/Tools/isac/Knowledge/Biegelinie.thy\#l174}.
}
in order to demonstrate the colouring helpful for programmers (compare the old 
bare string version in \cite[p.~92]{wn:proto-sys}).
\begin{figure} [htb]
  \centering
  \includegraphics[width=\textwidth]{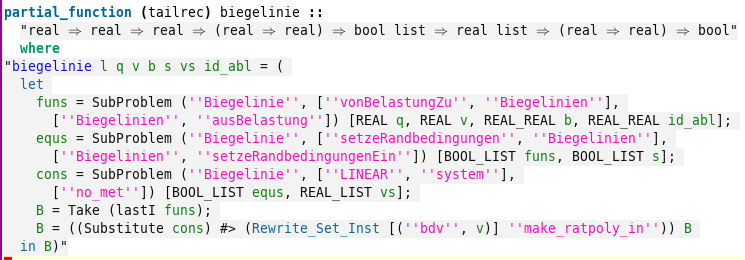}
  \caption{The program implemented by the function package}
  \label{fig:fun-pack-biegelinie}
\end{figure}
The program is implemented as a \texttt{\small partial\_function}: it calculates results for a whole class of problems and termination cannot be proven in such generality, although preconditions guard execution.
The program reflects a ``divide \& conquer'' strategy by calling three \texttt{\small SubProblem}s. It was implemented for field-tests in German-speaking countries, so arguments of the \texttt{\small SubProblem}s are in German, because they appear in the calculation as well. The third \texttt{\small SubProblem} adopts the format from Computer Algebra. The calculation, if finished successfully, looks as follows (a screen-shot of intermediate steps on the prototype's front-end is shown in \cite[p.~97]{wn:proto-sys}):
\label{biegel-calc}

{\footnotesize\begin{tabbing}
123\=123\=12\=12\=12\=12\=12\=123\=123\=123\=123\=\kill
\>01\> Problem (Biegelinie, [Biegelinien]) \\
\>02\>\> Specification: \\
\>03\>\> Solution: \\
\>04\>\>\> Problem (Biegelinie, [vonBelastungZu, Biegelinien]) \\
\>05\>\>\> $[$\>$V\,x = c + -1\cdot q_0\cdot x, $\\
\>06\>\>\>    \>$M_b\,x = c_2 + c\cdot x + \frac{-1\cdot q_0}{2\cdot x ^ 2}, $\\
\>07\>\>\>    \>$\frac{d}{dx}y\,x =  c_3 + \frac{-1}{EI} \cdot (c_2\cdot x + \frac{c}{2\cdot x ^ 2} + \frac{-1\cdot q_0}{6\cdot x ^ 3}, $\\
\>08\>\>\>    \>$y\,x =  c_4 + c_3\cdot x +  \frac{-1}{EI} \cdot  (\frac{c_2}{2}\cdot x ^ 2 + \frac{c}{6}\cdot x ^ 3 + \frac{-1\cdot q_0}{24}\cdot x ^ 4)\;\;]$ \\
\>09\>\>\> Problem (Biegelinie, [setzeRandbedingungen, Biegelinien])\\
\>10\>\>\> $[$\>$L \cdot q_0 = c,\; 0 = \frac{2 \cdot c_2 + 2 \cdot L \cdot c + -1 \dot L ^ 2 \cdot q_0}{2},\; 0 = c_4,\; 0 = c_3\;\;]$\\
\>11\>\>\> solveSystem $(L \cdot q_0 = c,\; 0 = \frac{2 \cdot c_2 + 2 \cdot L \cdot c + -1 \cdot L ^ 2 \cdot q_0}{2},\; 0 = c_4,\; 0 = c_3\;\;],\; [c, c_2, c_3, c_4])$\\
\>12\>\>\> $[$\>$c = L \cdot q_0 ,\; c_2 = \frac{-1 \cdot L ^ 2 \cdot q_0}{2},\; c_3 = 0,\; c_4 = 0]$\\
\>13  \` Take $y\,x =  c_4 + c_3\cdot x +  \frac{-1}{EI} \cdot  (\frac{c_2}{2}\cdot x ^ 2 + \frac{c}{6}\cdot x ^ 3 + \frac{-1\cdot q_0}{24}\cdot x ^ 4)$ \\

\>14\>\>\> $y\,x = c_4 + c_3 \cdot x + \frac{-1}{EI} \cdot (\frac{c_2}{2} \cdot x ^ 2 + \frac{c}{6} \cdot x ^ 3 + \frac{-1 \cdot q_0}{24} \cdot x ^ 4)$\\
\>15  \` Substitute $[c,c_2,c_3,c_4]$ \\

\>16\>\>\> $y\,x = 0 + 0 \cdot x + \frac{-1}{EI} \cdot (\frac{\frac{-1 \cdot L ^ 2 \cdot q_0}{2}}{2} \cdot x ^ 2 + \frac{L \cdot q_0}{6} \cdot x ^ 3 + \frac{-1 \cdot q_0}{24} \cdot x ^ 4)$\label{exp-biegel-Substitute}\\
\>17  \` Rewrite\_Set\_Inst $([({\it bdv},x)], {\it make\_ratpoly\_in})$ \\

\>18\>\>\> $y\;x = \frac{q_0 \cdot L ^ 2}{4 \cdot EI} \cdot x ^ 2 + \frac{L \cdot q_0 }{6 \cdot EI} \cdot x ^ 3 + \frac{q_0}{24 \cdot EI} \cdot x ^ 4$\\
\>19\> $y\;x = \frac{q_0 \cdot L ^ 2}{4 \cdot EI} \cdot x ^ 2 + \frac{L \cdot q_0 }{6 \cdot EI} \cdot x ^ 3 + \frac{q_0}{24 \cdot EI} \cdot x ^ 4$
\end{tabbing}}
The calculation is what a student interacts with. Above the \texttt{\small Specification} is folded in, the specification phase and respective user interaction is skipped here, because user-interaction in this phase is not guided by \LI{} (while the result of the phase, a concise formal specification of the problem is a necessary pre-condition for \LI). The \texttt{\small Solution} must be constructed step-wise line by line (while the line numbers do not belong to the calculation) in forward reasoning.
\label{1-2-3}A student inputs either (1) a term (displayed on the left above with indentations) or (2) a tactic (shifted to the right margin). If the student gets stuck (3) a next step (as a term or a tactic) is suggested by \LI{}, which works behind the scenes; the corresponding functions have been introduced in \S\ref{concept-LI} (for (1) \texttt{\small locate\_input\_term}, (2) \texttt{\small locate\_input\_tactic} and for (3) \texttt{\small find\_next\_step}). A student can combine actions (1)..(3) arbitrarily.

\medskip
This example hopefully clarifies the relation between program and calculation in \LI{}: execution of the program stops at tactics (above called  break-points), displays the tactic or the result produced by the tactic (as decided by the dialogue-module) in the calculation under construction, hands over control to the student working on the calculation and resumes checking user input.

\paragraph{A typical example from Computer Algebra}\label{expl-rewrite} shall shed light on a notions introduced later, on tacticals. The program in Fig.\ref{fig:fun-pack-diff} 
simplifies rational terms (typed as ``real'' due to limited development resources):
\begin{figure} [htb]
  \centering
  \includegraphics[width=0.95\textwidth]{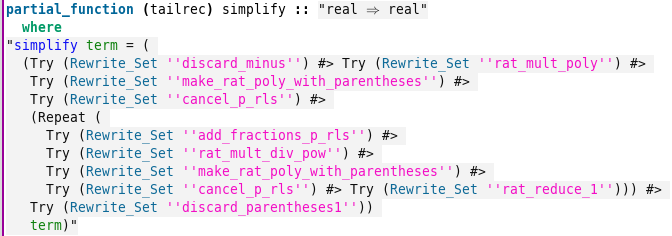}
  \caption{Simplification implemented by the function package}
  \label{fig:fun-pack-diff}
\end{figure}

The program executes the tactic \texttt{\small Rewrite'\_Set}\footnote{Isabelle's document preparation system preserves the apostroph from the definition, while Isabelle's pretty printer drops it in the presentation of the program as above. The apostroph excapes the underscore as required by Isabelle's lexer.}
with different canonical term rewriting systems (called ``rule sets'' in \sisac) guided by tacticals. Chained functions (by \texttt{\small \#>}) 
are applied curried to \texttt{\small term}. The initial part of the chain, from \texttt{\small ''discard\_minus''} to \texttt{\small ''cancel''} does preprocessing, for instance replaces $\;a - b\;$ by $\;a + (-b)\;$ in \texttt{\small ''discard\_minus''} for reducing 
the number of theorems to be used in simplification.
The second chain of \texttt{\small Rewrite\_Set} is \texttt{\small Repeat}ed until no further rewrites are possible; this is necessary in case of nested fractions.

In cases like the one above, a confluent and terminating term rewrite system, not only \emph{all} correct input terms are accepted, but also \emph{all} incorrect input is rejected as \texttt{\small Not\_Derivable} by the function  \texttt{\small locate\_input\_term}.

\subsection{\LI's Relevance for Educational Mathematics Software}\label{edu}
Main concern of this paper is describing technical aspects of \LI. But here is a brief summary of promises for education, which should make the paper selfcontained for educators --- promises derived from \LI's three functions introduced on p.\pageref{sig-lucin} and interpreted with respect to students' interaction on p.\pageref{1-2-3}.

\paragraph{\LI{} is generally usable} by all software, which constructs a solution for a given problem --- this, in principle, includes all programs, for instance, written in Mathematica \footnote{\url{https://www.wolfram.com/mathematica/online/}} and millions of educational software products: just (re-)write the programs calculating a solution as shown in Fig.\ref{fig:fun-pack-biegelinie}. Doing so would cover most of mathematics as taught at engineering faculties as well as taught at high-schools.

Of course, that is an endless deal of work: but it can be done by lecturers and teachers, who like writing software for the problems they teach. And, of course, in order to make their development efficient, comprehensive libraries need to be available free of charge (like some of the libraries available for Mathematica). Such libraries would be concern of publicly funded academic development projects.

Such projects can build upon mechanised mathematics knowledge, which goes far beyond the contents of undergraduate studies (for instance in Isabelle's archive of formal proofs~\cite{url-afp}) in mathematics and in engineering studies. From a certain point, development would become efficient and sustainable, because based on software with a huge user community extending the communities working on mechanisation of mathematics knowledge.

\paragraph{\LI{} features ``models of mathematics'',} transparent, complete and interactive models, as already described in \cite{EPTCS290.6} and briefly repeated here. Mathematical problem solutions (in the sequel called \emph{calculations}) can be ``transparent'' by mouse-click on an element of a calculations, which immediately leads to the respective definition~\footnote{This feature
  is known from software development environments (IDEs). Since Isabelle/jEdit realises this feature on terms~\cite{DBLP:journals/corr/Wenzel14}, it appears an indispensable requirement for mathematics software at the state of the art.
} and related properties together with (more or less) readable proofs. Thus mathematical calculations can be ``complete'' with respect to deduction from first principles (where students can follow these deductions at their own pace, or even not). And mathematical calculations can be ``interactive'' relying on \LI{} as described in this paper.

What makes such calculations to ``models of mathematics'' is the fact, that it covers all of mathematics which can be mechanised, while most computer mathematicians agree that (almost all of) mathematics can be mechanised (as soon as it is fully understood). And  each item of knowledge and each step in a calculations is proven correct, unambiguously and based on logic --- which fullfills the main property of mathematics, distinguishing it from other sciences. As, in history of science, rigorous logic was important to overcome belief and magic, comprehending this main property is an important goal for education in mathematics and in science. However, educational practice shows, that this goal is hard to achieve.

The educational gain from mechanised ``models of mathematics'' can be seen as follows: Many students \emph{prefer explanations in human language over looking at the structure of formulas} --- where the latter is indispensable for developing confidence in mathematics and for overcoming the impression of authority or magic eventually. Software works mechanically on formulas (and bypasses natural language). Thus it continuously offers direct experience with the mechanical nature of mathematics (the science of mechanisation of thinking~\cite{RISC4777}) --- at any time a student is ready to recognise it, a steady mechanical offer to mature the understanding of formal language. Such understanding wouldmake academic lectures on formal logic, on meta-mathematics, on theory of science and the like better comprehensible. And, probably, freshmen would start with more confidence (and in larger numbers) into science studies.

\paragraph{\LI{} features ``systems that explain themselves''} as already addressed in \cite{wn:cme-ei-18,EPTCS290.6} and increases the range of learning scenarious to be supported by software: from high-school up to elderly, who want to understand ``what mathematics really is'', from organised learning at various institutions up to individual study in private settings, from independent learning via learning by trial and error via ``flipped classroom instruction''~\cite{EMS-math-ethics}, private teaching and exercising difficult problem solutions up to written exams with automated evaluation.

The crucial feature for systems that explain themselves in mathematics is \texttt{find\_next\_step} as introduced on p.\pageref{sig-lucin}: a student always can request a hint from the system for a next step towards a solution --- and this feature also makes clear, that flexible limitation is needed to adapt to realistic learning scenarios. Such adpation is concern of a dialogue-module: by use of pattern-matching on the next step a hint can be given in full or partially (from a formula or a tactic, see the example calculation on p.\pageref{biegel-calc}), can be a list of options or can be idle (in case of a written exam).

Such a dialogue-module is implemented as a stub in \sisac's prototype~\cite{wn:proto-sys}, but several field tests~\cite{imst-htl06,imst-htl07,imst-hpts08} already clarified a lot of requirements for such a module; realisation shall be in collaboration with experts in educational psychology --- which exhibits a great advantage of \LI{} over present principles of educational software: design and development in computer mathematics (writing \LI-programs, mechanising deductive knowledge) are strictly separated from design and development of dialogues (extending prototypes like~\cite{mkienl-bakk} and adopting educational theories).

\section{Adaptation of Isabelle's Functions}\label{lucin-funpack}
Isabelle's function package presents functions in ``inner syntax'' to users, i.e. as terms in Isabelle/HOL. The \LI{} design recognised these terms suitable for parse trees of programs, Isabelle realised the same idea in a more general way with the function package a few years later --- so migration from \sisac's programs to Isabelle's functions was surprisingly easy. The main features required were tactics, tacticals and program expressions as described below.

\subsection{Tactics Creating Steps in Calculations}\label{tactics}
The examples in the previous section showed how tactics in programs create steps in calculations. Tactics in programs are defined as follows\footnote{
\url{https://hg.risc.uni-linz.ac.at/wneuper/isa/file/df1b56b0d2a2/src/Tools/isac/ProgLang/Prog_Tac.thy\#l36}}
:%
\begin{verbatim}
  consts
    Calculate          :: "[char list, 'a] => 'a"
    Rewrite            :: "[char list, 'a] => 'a"
    Rewrite'_Inst      :: "[(char list * 'a) list, char list, 'a] => 'a"   
    Rewrite'_Set       :: "[char list, 'a] => 'a"
    Rewrite'_Set'_Inst :: "[((char list) * 'a) list, char list, 'a] => 'a"
    Or'_to'_List       :: "bool => 'a list"
    SubProblem         ::
      "[char list * char list list * char list list, arg list] => 'a"
    Substitute         :: "[bool list, 'a] => 'a"
    Take               :: "'a => 'a"
\end{verbatim}

These tactics transform a term of some type \texttt{\small 'a} to a resulting term of the same type.  \texttt{\small Calculate} applies operators like $+,-,*$ as \texttt{\small char\ list} (i.e. a string in inner syntax) to numerals of type  \texttt{\small 'a}. The tactics beginning with \texttt{\small Rewrite} do exactly what they indicate by applying a theorem (in the program given by type \texttt{\small char\ list}) or a list of theorems, called  ``rule-set''. The corresponding \texttt{\small \_Inst} variants instantiate bound variables with respective constants before rewriting (this is due to the user requirement, that terms in calculations are close to traditional notation, which excludes $\lambda$-terms), for instance modelling bound variables in equation solving. \texttt{\small Or\_to\_List} is due to a similar requirement: logically appropriate for describing solution sets in equation solving are equalities connected by $\land,\lor$, but traditional notation uses sets (and these are still lists for convenience). \texttt{\small SubProblem}s not only take arguments (\texttt{\small arg\_list} like any (sub-)program, but also three references into \sisac's knowledge base  (theory, formal specification, method) for guided interaction in the specification phase.

Tactics appear simple: they operate on terms adhering to one type --- different types are handled by different tactics and (sub-) programs; and they cover only basic functionality --- but they operate on terms, which are well tooled by Isabelle and which can contain functions evaluated as program expressions (which will be introduced in \S\ref{prog-expr} below).

Section \S\ref{expl-lucin} showed how tactics can be input by students. So tactics in programs have analogies for user input with type \texttt{\small Tactic.input} defined at\footnote{
\url{https://hg.risc.uni-linz.ac.at/wneuper/isa/file/df1b56b0d2a2/src/Tools/isac/MathEngBasic/tactic-def.sml\#l122}}. These tactics also cover the specification phase (wich is out of scope of the paper). And there is another \texttt{\small Tactic.T} defined at\footnote{
\url{https://hg.risc.uni-linz.ac.at/wneuper/isa/file/df1b56b0d2a2/src/Tools/isac/MathEngBasic/tactic-def.sml\#l241}} for internal use by the mathematics-engine, which already appeared in the signature of Lucas-Interpretation on p.\pageref{sig-lucin}.

\subsection{Tacticals Guiding Flow of Execution}\label{tacticals}
The example on p.\pageref{expl-rewrite} for canonical rewriting showed, how tacticals guide the flow of execution. The complete list of tacticals is as follows.%
\begin{verbatim}
  consts
    Chain    :: "['a => 'a, 'a => 'a, 'a] => 'a" (infixr "#>" 10)
    If       :: "[bool, 'a => 'a, 'a => 'a, 'a] => 'a"
    Or       :: "['a => 'a, 'a => 'a, 'a] => 'a" (infixr "Or" 10)
    Repeat   :: "['a => 'a, 'a] => 'a" 
    Try      :: "['a => 'a, 'a] => 'a"
    While    :: "[bool, 'a => 'a, 'a] => 'a" ("((While (_) Do)//(_))" 9)
\end{verbatim}
\texttt{\small Chain}  is forward application of functions and made an infix operator \#$>$ 
; \texttt{\small If} decides by \texttt{\small bool}ean expression for execution of one of two arguments of type \texttt{\small 'a\ $Rightarrow$ 'a} (which can be combinations of tacticals or tactics); \texttt{\small Or} decides on execution of one of the arguments depending of applicability of tactics; \texttt{\small Repeat} is a one way loop which terminates, if the argument is not applicable (e.g. applicability of a theorem for rewriting on a term) any more; \texttt{\small Try} skips the argument if not applicable and \texttt{\small While} is a zero way loop as usual.

\subsection{Program Expressions to be Evaluated}\label{prog-expr}
Some tacticals, \texttt{\small If} and \texttt{\small While}, involve boolean expressions, which need to be evaluated: such expressions denote another element of programs. This kind of element has been shown in the example on p.\pageref{fig:fun-pack-biegelinie} as argument of  \texttt{\small Take}: sometimes it is necessary to pick parts of elements of a calculation, for instance the last element from a list. So Isabelle's \texttt{\small List} is adapted for \sisac's purposes in \texttt{\small ListC}\footnote{
\url{https://hg.risc.uni-linz.ac.at/wneuper/isa/file/df1b56b0d2a2/src/Tools/isac/ProgLang/ListC.thy}}.
Such expressions are substituted from the environment in \texttt{\small Istate.T}, evaluated to terms by rewriting (and for that purpose using the same rewrite engine as for the tactics \texttt{\small Rewrite}*) and marked by the constructor \texttt{\small Term\_Val} which is introduced below.

\section{Implementation of \LI}\label{LI-impl}
The implementation of the interpreter is as experimental as is the respective programming language introduced in the previous section. So below there will be particularly such implementation details, which are required for discussing open design \& implementation issues.

All of the function package's syntactic part plus semantic markup is perfect for \LI. The evaluation part of the function package, however, implements automated evaluation in one go and automated code-generation~\cite{code-gen-tutorial} --- both goals are not compatible with \sisac's goal to feature step-wise construction of calculations, so this part had to be done from scratch.

\subsection{Scanning the Parse Tree}\label{scanning}
Isabelle's function package parses the program body from function definitions to terms, the data structure of simply typed $\lambda$ terms, which also encode the objects of proofs. Thus there is a remarkable collection of tools, readily available in Isabelle; but this collection does not accommodate the requirement of scanning a term to a certain location, remembering the location and returning there later, as required by \LI. So this has been introduced

\begin{verbatim}
  datatype lrd = L | R | D
  type path = lrd list
  
  fun at_location [] t = t
    | at_location (D :: p) (Abs(_, _, body)) = at_location p body
    | at_location (L :: p) (t1 $ _) = at_location p t1
    | at_location (R :: p) (_ $ t2) = at_location p t2
    | at_location l t =
      raise TERM ("at_location: no " ^ string_of_path l ^ " for ", [t]);
\end{verbatim}

\noindent with a \texttt{\small path} to a location according to the term constructors \texttt{\small \$} and \texttt{\small Abs}. This is an implementation detail; an abstract, denotational view starts with this datatype%

\begin{verbatim}
  datatype expr_val =
      Term_Val of term  | Reject_Tac
    | Accept_Tac of Istate.pstate * Proof.context * Tactic.T
\end{verbatim}

\noindent which models the meaning of an expression in the parse-tree: this is either a  \texttt{\small Term\_Val}ue  (introduced in \S\ref{prog-expr}) or a tactic (introduced in \S\ref{tactics}); the latter is either accepted by  \texttt{\small Accept\_Tac} or rejected by  \texttt{\small Reject\_Tac}; an error value is still missing.
The arguments of the above constant \texttt{\small Accept\_Tac} are the same as introduced for \LI{} in \S\ref{concept-LI}. Thus \texttt{\small expr{\char`\_}val} is the return value for functions scanning the parse-tree for an acceptable tactic:%

\begin{verbatim}
  scan_to_tactic: term * (Calc.T * Proof.context) -> Istate.T -> expr_val;
  scan_dn: Calc.T * Proof.context -> Istate.pstate -> term -> expr_val;
  go_scan_up: term * (Calc.T * Proof.context) -> Istate.pstate -> expr_val;
  scan_up: term * (Calc.T * Proof.context) -> Istate.pstate -> term -> expr_val;
\end{verbatim}

\noindent The first argument of each function above, if of type  \texttt{\small term}, is the whole program required for scanning given a particular \texttt{\small path} using \texttt{\small at\_location}. This is not required by \texttt{\small scan\_dn}, which is simple top-down scanning. The rightmost argument, if of type  \texttt{\small term}, serves matching in \texttt{\small scan\_dn} and in  \texttt{\small scan\_up}~\footnote{
The code for these functions is found at \url{https://hg.risc.uni-linz.ac.at/wneuper/isa/file/113997e55e71/src/Tools/isac/Interpret/lucas-interpreter.sml\#l125} and at \url{https://hg.risc.uni-linz.ac.at/wneuper/isa/file/113997e55e71/src/Tools/isac/Interpret/lucas-interpreter.sml\#l192} respectively.}

The central function is \texttt{\small scan\_to\_tactic} which accomplishes what the identifier indicates: if a student creates a next step at a current position in the \texttt{\small Solution} of a problem, then this position is associated with a certain \texttt{\small path} in the parse-tree, see example on p.\pageref{biegel-calc}. \LI{} supports creating steps by one of the three actions described on p.\pageref{1-2-3}. According to the action chosen \LI{} scans from the certain \texttt{\small path} to the next tactic found in the parse-tree in order to either check the input formula (1) or tactic (2), or to propose a next step (3).
Function \texttt{\small scan\_to\_tactic} calls either \texttt{\small scan\_dn}, if interpretation has just started a new program (\texttt{\small path = []})
or \texttt{\small scan\_dn} via \texttt{\small go\_scan\_up}:%

\begin{verbatim}
  fun scan_to_tactic (prog, cc) (Pstate (ist as {path, ...})) =
      if path = []
      then scan_dn cc (ist |> set_path [R]) (Program.body_of prog)
      else go_scan_up (prog, cc) ist
    | scan_to_tactic _ _ =
      raise ERROR "scan_to_tactic1: uncovered pattern in fun.def"
\end{verbatim}

\noindent\texttt{\small scan\_dn} sets the path in the interpreter state \texttt{\small ist} to \texttt{\small R}, the program body, while \texttt{\small go\_scan\_up} goes one level up the path in order to call \texttt{\small scan\_up}. The latter uses the \texttt{\small path} found in \texttt{\small ist}, the interpreter state from the previous step.

It might be helpful to state, that the functions \texttt{\small scan\_dn} and \texttt{\small scan\_up} need not be mutually recursive: as soon as these functions have found their goal, an appropriate tactic, \LI{} terminates, stores the current state and hands over control to the student. Afterwards, depending on the \texttt{\small path}, either \texttt{\small scan\_dn} or \texttt{\small scan\_up} are called, so they are independent from each other.%

\subsection{Use of Isabelle's Contexts}\label{ctxt}
Isabelle's \textit{``logical context represents the background that is required for formulating statements and composing proofs. It acts as a medium to produce formal content, depending on earlier material (declarations, results etc.).''} \cite{implem-tutorial}. The \sisac-prototype introduced Isabelle's \texttt{\small Context} in collaboration with a student~\cite{mlehnf:bakk-11}, now uses them throughout construction of problem solutions and implements a specific structure \texttt{\small ContextC}\footnote{A closing ``C'' indicates an \sisac{} extension, see {\url{https://hg.risc.uni-linz.ac.at/wneuper/isa/file/00612574cbfd/src/Tools/isac/CalcElements/contextC.sml}}}.

At the beginning of the specification phase (briefly touched by tactic \texttt{\small SubProblem} in \S\ref{expl-lucin} and explained in little more detail in the subsequent section) an Isabelle theory must be specified, this is followed by \texttt{\small Proof\_Context.init\_global} in the specification module. Then \texttt{\small Proof\_Context.initialise'} takes a formalisation of the problem as strings, which are parsed by \texttt{\small  Syntax.read\_term} using the context, and finally the resulting term's types are recorded in the context by \texttt{\small Variable.declare\_constraints}. The latter relieves students from type constraints for input terms. Finally, as soon as a problem type is specified, the respective preconditions are stored by \texttt{\small ContextC.insert\_assumptions}.

\medskip
During \LI{} the context is updated with assumptions generated by conditional rewriting and by switching to and from sub-programs. An example for the former is tactic  \texttt{\small SubProblem} applied with theorem $x \ne 0 \Rightarrow (\frac{a}{x} = b) = (a = b\cdot x)$ during equation solving.

A still not completely solved issue is switching to and from \texttt{\small SubProblem}s; the scope of an interpreter's environment is different from a logical context's scope. When calling a sub-program, \sisac{} uses \texttt{\small Proof\_Context.initialise}, but returning execution from a \texttt{\small SubProblem} is not so clear. For instance, if such a sub-program determined the solutions $[x=0,\;x=\sqrt{2}]$, while the calling program maintains the assumption $x\not=0$ from above, then the solution $x=0$ must be dropped, this is clear. But how determine in full generality, which context data to consider when returning execution to a calling program? Presently the decision in \texttt{\small ContextC.subpbl\_to\_caller}\footnote{
\url{https://hg.risc.uni-linz.ac.at/wneuper/isa/file/ce071aa3eae4/src/Tools/isac/CalcElements/contextC.sml\#l73}} is, to transfer all content data which contain at least one variable of the calling program and drop them on contradiction.

\subsection{Guarding and Embedding Execution}\label{embedding}
\texttt{\small partial\_function}s as used by \LI{} are alien to HOL for fundamental reasons. And when the \sisac-project started with the aim to support learning mathematics as taught at engineering faculties (in that generality, see \S\ref{edu}), it was clear, that formal specifications should guard execution of programs under \LI{} (i.e. execution only starts, when the specification's preconditions evaluate to true).

So formal specification is required for technical reasons \emph{and} for educational reasons in engineering education. The \sisac-project designed a separate specification phase, where input to a problem and corresponding output as well as preconditions and post-condition are handled explicitly by students; example in  Fig.\ref{fig:fun-pack-biegelinie} shows several \texttt{\small SubProblem}s, which lead to mutual recursion between specification phase and phases creating a solution (the latter supported by \LI). Embedding \LI{} into a dialogue-module is required in order to meet user requirements as already discussed in \S\ref{edu}.

\section{Lessons Learned ...}\label{learned}
As already mentioned, development of Isabelle and development of \LI{} went in parallel for a long time --- a great opportunity for learning in the \sisac-project.

\subsection{... from the Isabelle Project}\label{learned-devel}
``\emph{Isabelle was not designed; it evolved. Not everyone likes this idea}'' 
said Lawrence C. Paulson in ``Isabelle: The Next 700 Theorem Provers''~\cite{paulson700}.

When the \sisac-project started about the year 2000, Isabelle's code structure still reflected the enormous efforts of Paulson to make a great idea a usable product. Such a situation naturally leads to code, where placement of code is determined by the context of new feature requests. In the meanwhile Isabelle evolved to a didactic model in functional programming at a large scale: polymorphic higher order functions take complex function-arguments, which allow to postpone type definitions to locations according to functionality; and such functionality is distilled to small abstract structures. Together with canonical argument order, function combinators and canonical iteration  \cite[p.~15-17]{implem-tutorial} this gives elegant code with almost no glue, so the hint ``never copy \& paste a piece of code'' could disappear from the implementation manual some time ago. And now a layered structure becomes apparent, best reflected by exception hierarchies. So Isabelle is in a state, where ``everything can be changed anywhere in the code'' in research\&development.

Also Isabelle's development process is a didactic model in efficient collaborative development distributed all around the world and in minimisation of administrative efforts. Visible outcomes from this process are formally checked documentation and a code repository with minimal change-sets, which denote essential feature changes with a minimum of updated code.

\medskip
The \sisac-project started in a situation, where software and user requirements might have been as unclear as with early Isabelle, albeit on another level. \sisac's present code structure still looks much more like (very!) early Isabelle than present Isabelle. This paper was the occasion to make \LI-related components in \sisac{} as close to Isabelle's style --- and it is a hard experience, that this cannot be done in one go and will require many ``rounds of reform'': \sisac's code structure, even around \LI, is still far off Isabelle's quality. This is the same with minimal change-sets: in the present state of \sisac's code so much improvements occur along current work, that it is just inefficient to separate respective change-sets in many cases.

\subsection{... from Isabelle's Function Package}\label{learned-feature}
When the development of \sisac{} started, a glance at Isabelle's front-end convinced everyone, that educational software could \emph{not} use it. Now, two decades later, it appears clear, that \sisac{} is best advised to re-use Isabelle/JEdit based on Isabelle/PIDE, the integrated proof development environment as the state of the art.
This is particularly evident from the work presented in this paper, shifting Lucas-Interpretation into Isabelle/Isar's function definition, which makes all the advanced features of Isabelle/JEdit available for the working programmer:
\begin{itemize}
\item \textbf{Syntax errors} are indicated accurately at the right location; finding errors in programs represented as strings was a nightmare, if programs comprised more than a couple lines of code.
\item \textbf{Type annotations} disappear from the program and sidestep to the heading signature; the result is much better readable.
\item \textbf{Syntax highlighting} indicates how identifiers are interpreted, as constants, as free variable, as strings, etc --- very instructive for working programmers.
\item \textbf{Free variables} on the right-hand-side of assignments are rejected by the function package, while these were accepted by term parsing.
\item \textbf{Semantic annotations} support the programmer, in particular the tooltip popups triggered by hovering and clicking with the mouse, see~\cite[p.~30]{isabelle-jedit}.
\end{itemize}
These features came for free, when \sisac's programs were shifted into Isabelle/Isar's function definition --- and were immediately fruitful: Not only the implementation of programs is much more efficient, also errors have been revealed. The features helped to detect additional free variables on the right-hand side in programs (see for instance this kind of errors in \cite[p.~92]{wn:proto-sys}) and triggered improved handling of program arguments in \LI.

\subsection{... for Further \isac-Development}\label{learned-isac}
What \sisac{} can learn from Isabelle's code structure and from Isabelle's development process has been described in \S\ref{learned-devel}, while Isabelle's packaging and deployment has not yet been mentioned, \sisac{} would benefit as well.

Future development in \sisac{} will follow a list of steps as announced in \cite[p.~102-103]{wn:proto-sys}; the work on \LI{} presented in this paper is the first respective step and confirms the relevance of the other steps. However, in the meanwhile difficulties in funding a corresponding project became apparent. Indeed, the many challenges already identified in~\cite{plmms10}, suggest intermediate steps on the way to educational software for engineering disciplines as aimed at in the \sisac-project.

A case study on GCD~\cite{wneuper:gcd-coimbra} already investigated possibilities for such intermediate steps: the Euclidean Algorithm creates a ``calculation'' as presented above, if the respective invariant (or fixpoint) is output at each recursive call of the algorithm --- in this example \LI{} could immediately be used to explain and to exercise calculation of the greatest common divisor of integers and polynomials. So the next challenge is to adapt \LI{} such that arbitrary functions can be executed step-wise --- as an experimental approach to study algorithms in the Archive of Formal Proofs~\footnote{\url{https://www.isa-afp.org/}}. Algorithms in formal logic like Resolution, Binary Constraint Propagation or the DPLL algorithms can be implemented for step-wise interpretation by \LI{} already at the present state.

Such intermediate steps postpone the requirement of 2-dimensional term representation on screen --- the requirement identified as most urgently missing in \sisac's field tests, and also not provided by Isabelle/jEdit: Isabelle's line-oriented presentation of terms perfectly accomplishes user requirements for proof assistants but \emph{not} for mathematics in general, where fractions are read as $\frac{a}{b}$ and not as $a/b$. However, Isabelle's semantic markup in the front-end with links to types, definitions etc is indispensable in software at the actual state of computer mathematics. Presently there is no formula editor available, which features both, 2-dimensional representation \emph{and} semantic markup --- a separate challenge for academic open source development: just add two integers to Isabelle's \texttt{\small Position}, re-use what Knuth has implemented in \LaTeX{} for an editor (probably embedded into Isabelle/jEdit) --- and create significant impact on mathematics software of various kinds (including Isabelle itself)!

\section{Summary and Conclusions}\label{summary-concl}
This paper gave the first technically concise description of Lucas-Interpretation (\LI) and showed how concepts from Automated Reasoning support flexible student interaction by reliable check of student input. The description focuses key points and gives many pointers into the code in a freely accessible repository\footnote{\url{https://hg.risc.uni-linz.ac.at/wneuper/isa/}}. This is to invite readers to re-use prototyped code and/or the concept of \LI; the previous section gave some hints for re-use and for future development, more will be thinkable after further experiences.

The example from structural engineering on p.\pageref{bend-line-expl} makes two practical consequences of \LI{} clear. Firstly, \emph{all} kinds of mathematical problems, which can be described by an algorithm, can be associated with automatically generated user-guidance using \LI{} -- this covers most of mathematics as taught at arbitrary engineering faculties. Secondly implementation of respective algorithms is simple, see Fig.\ref{fig:fun-pack-biegelinie}; and programming becomes more efficient with each  \texttt{\small SubProblem} implemented earlier: re-using these makes development sustainable.

\paragraph{Final conclusions} address education, since \LI{} has been invented for educational purposes and the \sisac-project tries to adopt and adapt technologies with pedagogical concepts in mind.

The migration of \LI{} to the function package appears to illustrate the flexible conception of the Isabelle framework, and the integrative nature of this conception confirms the hope to realise ``complete, transparent \& interactive models of mathematics''~\cite{EPTCS290.6} for education. Such mechanical models might provide experience with mathematics as the ``discipline in mechanisation of thinking''~(Bruno Buchberger~\cite{RISC4777}) 
as an indispensable complement to teaching mathematics with human intuition --- an experience required to understand not only the strengths, but also the limitations of mathematical thinking technology~\cite{EMS-math-ethics}.


\nocite{*}
\bibliographystyle{eptcs}
\bibliography{root}

\end{document}